\documentclass{appolb}
\usepackage{epsfig}
\newcommand{\pp}{\mbox{p--p}}
\newcommand{\PbPb}{\mbox{Pb--Pb}}
\newcommand{\lsim}{\,{\buildrel < \over {_\sim}}\,}

\newcommand{\av}[1]{\left\langle #1 \right\rangle}
\newcommand{\sqrts}{\sqrt{s}}
\newcommand{\sqrtsNN}{\sqrt{s_{\mathrm{NN}}}}
\newcommand{\dNchdeta}{{\rm d}N_{ch}/{\rm d}\eta}
\newcommand{\Npart}{N_{\rm part}}
\newcommand{\Ncoll}{N_{\rm coll}}
\newcommand{\mub}{\mathrm{\mu b}}
\newcommand{\pt}{p_{\rm t}}
\newcommand{\ptave}{\langle p_{\rm t} \rangle}
\newcommand{\Et}{E_{\rm t}}

\newcommand{\MeV}{\mathrm{MeV}}
\newcommand{\GeV}{\mathrm{GeV}}

\newcommand{\gev}{\mathrm{GeV}}
\newcommand{\tev}{\mathrm{TeV}}
\newcommand{\Raa}{R_{\rm AA}}
\newcommand{\Rcp}{R_{\rm CP}}
\newcommand{\Dzero}{{\rm D^0}}

\newcommand{\Dplus}{{\rm D^+}}
\newcommand{\jpsi}{{\rm J/}\psi}

\begin{document}
\title{ALICE Overview%
\thanks{Presented at Strangeness in Quark Matter 2011}%
}
\author{Francesco Prino (for the ALICE Collaboration)
\address{INFN, Sezione di Torino, Italy}
}
\maketitle
\begin{abstract}
A general overview of the results obtained 
by the ALICE experiment from the analysis of the
$\PbPb$ data sample collected at the end of 2010 
during the first heavy-ion run at the LHC is presented.
\end{abstract}
\PACS{25.75.-q}
  
\section{Introduction}

In November 2010 Pb nuclei were collided for the first time
at the LHC at the centre-of-mass energy $\sqrtsNN=2.76~\tev$,
about 14 times higher than the highest energy achieved at RHIC.
Three experiments (ALICE, ATLAS and CMS) collected data during the 5 weeks of 
running with Pb ions.
The integrated luminosity delivered by the LHC was 10~$\mub^{-1}$.

The study of $\PbPb$ collisions in the new energy regime attained at the LHC
is aimed at gaining deeper insight on the properties of nuclear 
matter at extreme conditions of temperature and energy density, where Lattice 
QCD predicts the matter to be in a Quark Gluon Plasma (QGP) state.
Experimental measurements at the LHC are a key benchmark for 
models that reproduce the features observed at lower collision energy.
Furthermore, results from the LHC are expected to address some of the issues 
that are not completely understood from the SPS and RHIC experiments (e.g. 
the $\jpsi$ suppression).
Finally, since the cross-section for QCD scatterings with high virtuality
increases steeply  with $\sqrts$, hard partons are abundantly produced
at the LHC, thus enabling high precision measurements for the experimental 
observables related to high momentum and heavy flavoured particles.

In the next sections, an overview of the ALICE results from the first 
heavy-ion run at the LHC is presented. 
%The observables from the ``soft'' physics sector are discussed in 
%sections~\ref{sec:glo},~\ref{sec:coll} and~\ref{sec:strang}, focusing on 
%particle multiplicities, particle correlations, collective motions and 
%strangeness production.
%The results from the ``hard'' probes (high momentum particles, heavy flavours 
%and quarkonia) are described in section~\ref{sec:hard}.

\section{Global event characteristics}
\label{sec:glo}

The multiplicity of produced particles, quantified by the charged particle 
density per unit of rapidity ($\dNchdeta$) at mid-rapidity, was measured as 
a function of the collision centrality~\cite{ALICEmult1, ALICEmult2}.
The measurement of the $\dNchdeta$ provides insight into the density of 
gluons in the initial stages and on the mechanisms of particle production.
The multiplicity in the 5\% most central collisions at the LHC is larger by a 
factor 2.2 with respect to central collisions at top RHIC energy.
The increase of multiplicity with centre-of-mass energy is 
steeper than the log$\sqrts$ trend observed at lower energies.
The centrality dependence of ($\dNchdeta$)/($\Npart$/2) has a similar shape 
to that observed at RHIC and is reasonably reproduced both by models based on
gluon saturation in the initial state and by two-component Monte Carlo 
models~\cite{ALICEmult2}.

The produced transverse energy $\Et$ was estimated by measuring 
the charged hadronic energy with the tracking system and adding the 
contribution of neutral particles.
The measured $\Et$ per pseudorapidity unit can be used to estimate the 
energy density with the Bjorken formula~\cite{Bjorken}.
%\begin{equation}
%\label{Bjorken}
%\varepsilon_{Bj}=\frac{1}{\mathcal{A} \tau} \left. \frac{d\Et}{dy}\right|_{y=0}
%\end{equation}
%where $\mathcal{A}$ is the transverse overlapping area of the colliding 
%nuclei and $\tau$ the formation time.
For the 5\% most central collisions at the LHC, the resulting value is 
$\varepsilon_{Bj}\tau \approx 15~\gev/({\rm fm}^2c)$ (where $\tau$ is the 
formation time), about a factor
3 larger than the corresponding one at RHIC~\cite{Plamen}.

The system size is measured from the HBT radii extracted from the study of 
two-pion correlations~\cite{ALICEhbt}.
For central collisions, the homogeneity volume is found to be larger 
by a factor two with respect to the one observed in central collisions at 
the top RHIC energy.
The decoupling time for mid-rapidity pions exceeds 10 fm/c and it is 
40\% larger than at RHIC.
The HBT radii were also extracted as a function of the event multiplicity
and compared with results at lower $\sqrts$.
$R_{long}$ is found to follow the linear trend with $\dNchdeta^{1/3}$ 
observed at lower energies.  
$R_{side}$ has a slightly different slope as a function of multiplicity
resulting to be at the lower edge of the uncertainty of the trend from 
lower $\sqrts$, while $R_{out}$ is clearly below the trend set by lower 
energies. 
This slower increase of $R_{out}$ with centre-of-mass energy can be explained 
in the framework of hydrodynamic models~\cite{Kisiel}.

%\begin{figure}[tb!]
%\centering
%\includegraphics[width=0.8\textwidth]{hbt-radii.pdf}
%\caption{HBT radii as a function of particle density as measured in heavy-ion 
%and $\pp$ collisions by ALICE compared with the results from lower energy 
%experiments. 
%Lines show linear fits to ALICE pp data and world AA data (except ALICE) 
%separately.}
%\label{fig:hbt}
%\end{figure}

\section{Collective motions}
\label{sec:coll}

The presence of collective motions arising from the large pressure gradients
generated by compressing and heating the nuclear matter is a typical
feature of the medium produced in heavy-ion collisions.
The radial flow, generated by the collective expansion of the 
fireball, is studied by measuring the transverse momentum ($\pt$) 
spectra of identified hadrons ($\pi$, K and p)~\cite{Preghenella,Guerzoni}.
The spectra reconstructed at the LHC are seen to be harder 
(i.e. characterized by a less steep distribution and a larger $\ptave$) than 
those measured at RHIC at $\sqrtsNN$=200~$\gev$.
This is a first indication for a stronger radial flow at the LHC.
The radial flow velocity at the thermal freeze-out, is estimated via
a blast-wave fit to the $\pi$, K and p spectra and, for the most central 
collisions at the LHC, it is found to be about 10\% higher than what 
observed in central collisions at top RHIC energy~\cite{Preghenella}.

The collective behaviour of the fireball is also studied from the
anisotropic flow patterns in the transverse plane that 
originate from the anisotropy in the spatial 
distribution of the nucleons participating in the collision.
Re-scatterings among the produced particles convert this initial
geometrical anisotropy into an observable momentum anisotropy.
Anisotropic flow is characterized by the Fourier coefficients 
$v_n=\langle\cos[n(\varphi-\Psi_n)]\rangle$, where $n$ is the 
order of the harmonic, $\varphi$ is the azimuthal angle of the
particle and $\Psi_n$ is the angle of the initial state spatial plane of 
symmetry.
The dominant harmonic is the elliptic flow, $v_2$, which is sensitive
to the properties of the system (equation of state, thermalization time,
viscosity) in the various stages of its evolution.
The $\pt$ integrated elliptic flow of charged particles at the 
LHC is found to increase by about 30\% from the highest RHIC energy of 
$\sqrts=200~\gev$~\cite{ALICEflow}.
The large value of elliptic flow indicates that the hot and dense matter 
created in heavy-ion collisions at LHC energies behaves like a strongly 
interacting fluid with exceptionally low viscosity, as already observed at 
RHIC.
The $\pt$-differential elliptic flow, $v_2(\pt)$, measured at the LHC is 
compatible with that observed at RHIC~\cite{ALICEflow}.
The 30\% increase in the $\pt$ integrated elliptic flow is therefore 
due to the increased average transverse momentum of the produced 
particles as a consequence of the stronger radial flow.
The larger radial flow leads also to a more pronounced dependence
of $v_2$ on the particle mass.
The mass splitting among $v_2(\pt)$ of identified pions, kaons and protons
at the LHC is actually found to be slightly but significantly larger than that
observed at lower collision energies~\cite{Noferini}. 

The predictions from hydrodynamics, based on the assumption that the QGP shear 
viscosity over entropy ratio does not change from RHIC to LHC~\cite{Heinz}, 
provide a good description of the $\pt$ spectra of pions and kaons in the
most central collisions, but disagree with the measured (anti-)proton spectrum,
both in shape and yield.
This hydrodynamic description reproduces well also the measured $v_2(\pt)$ 
for the three particle species for semi-peripheral (40--50\%) collisions, 
while for more central (10--20\%) reactions it misses the 
(anti-)protons~\cite{Noferini}.
An hybrid approach~\cite{HeinzVishnu} that couples viscous hydrodynamics of the 
QGP to a microscopic kinetic description of the hadronic phase reproduces 
significantly better both the $v_2(\pt)$ and the shape of the $\pt$ spectra of 
(anti-)protons~\cite{Preghenella,Noferini}.
This is an indication for a significant contribution from the extra flow 
built up in the hadronic phase.

Higher Fourier harmonics in the particle azimuthal distributions 
were also measured~\cite{ALICEhighharm}.
In particular, odd harmonics, e.g. $v_3$ and $v_5$, can take non-zero
values due to fluctuations in the spatial 
distribution of the participant nucleons, which cause
event-by-event fluctuations of the plane of symmetry $\Psi_n$ relative to the 
reaction plane.
According to hydrodynamics, odd harmonics are particularly sensitive to 
both the viscosity and the initial conditions of the system.
Indeed, the measured triangular flow ($v_3$) is significantly larger than
zero and does not depend strongly on centrality~\cite{youzhou}.
The presence of higher harmonics provides a natural explanation of the 
structures observed in the two-particle azimuthal correlations at low $\pt$, 
namely the ``ridge'' at $\Delta\varphi\approx0$ and large $\Delta\eta$ and the 
double-bump in the away side ($\Delta\varphi\sim\pi$), already observed at 
RHIC~\cite{RHICridge}. 
These structures in the two-particle correlations were studied in 
detail by performing a Fourier analysis of the $\Delta\eta - \Delta\varphi$ 
correlations with large $\Delta \eta$ gap~\cite{ALICE2part}.
The data are found to be well described by the first five terms of the Fourier 
series.
At low $\pt$ (i.e. $\pt \lsim 3-4~\gev/c$), the Fourier components 
extracted from two-particle correlations factorize into single-particle 
harmonic coefficients which agree with the measured anisotropic flow 
coefficients.
This indicates that the features observed in two-particle correlations at low 
$\pt$ are consistent with the collective response of the system to the 
initial state geometrical anisotropy.

\section{Particle abundances and strangeness production}
\label{sec:strang}

The $\pt$ differential distributions and the total yields 
were measured by ALICE for many hadronic species~\cite{Preghenella,Hippolyte}.
Charged pions, kaons, protons are identified via their 
dE/dx and time-of-flight. 
$K^0_S$ mesons, $\Lambda$, $\Xi$ and $\Omega$ hyperons are reconstructed 
from their decay topologies~\cite{Kalweit,Kalinak,Nicassio}.

The baryon/meson ratio, which is sensitive to the hadronization mechanism and 
in particular to quark recombination at the phase boundary,
is studied in the strangeness sector via the ratio $\Lambda$/$K^0_S$ as a 
function of $\pt$ in different centrality intervals.
The $\Lambda$/$K^0_S$ in peripheral (80-90\%) collisions stays below 0.7 and is
quite similar to what observed in $\pp$.
With increasing centrality the baryon/meson ratio increases 
developing a maximum at $\pt \approx 3~\GeV/c$ and it reaches a value 
$\Lambda$/$K^0_S \approx 1.5$ for the 5\% most central events.
With respect to what observed at lower energies, 
the baryon enhancement results larger at the LHC and the position in $\pt$ of 
the maximum of the $\Lambda$/$K^0_S$ is slightly shifted towards higher 
transverse momenta~\cite{Kalinak}.

The yields of pions, kaons, protons, $\Xi$ and $\Omega$ 
are extracted by integrating the measured $\pt$ spectra after fitting them
with a blast-wave function.
The particle abundances normalized to that of charged pions, can be compared 
with the predictions of a thermal model based on gran-canonical ensemble with 
temperature $T_{chem}=164~\MeV$ at the chemical freeze-out and baryochemical 
potential $\mu_{b}=1~\MeV$~\cite{Andronic}.
All the measured yields, with the notably exception of protons, are found to
agree with the thermal model predictions.
The measured proton/pion ratio falls significantly ($\approx 50\%$) below the 
thermal model expectation~\cite{Kalweit}.

The production of multi-strange baryons in $\PbPb$ collisions was also
compared with that measured in $\pp$ collisions.
This was done using the ratio between the yield of $\Xi$ and $\Omega$ hyperons
in $\PbPb$ and $\pp$ after normalizing to the number of participant nucleons. 
An enhancement of the production of $\Xi$ and $\Omega$ in heavy-ion collisions
with respect to $\pp$ is observed also at LHC energies~\cite{Nicassio}.
The enhancement is lower than that observed at SPS and RHIC 
energies, confirming its decreasing trend with increasing $\sqrts$.
This is a consequence of the smaller effect of canonical 
suppression for strangeness production in $\pp$ reactions at higher 
collision energies.

\section{Characterization of the medium with hard probes}
\label{sec:hard}

Particles with large transverse momentum and/or mass, which are produced  
in large-virtuality parton scatterings in the early stages of the collision, 
are powerful tools to probe the medium created in heavy-ion collisions.
Their production in nuclear collisions is expected 
to scale with the number of nucleon--nucleon collisions occurring in the 
nucleus--nucleus collision (binary scaling).
The experimental observable used to verify the binary scaling is the nuclear 
modification factor, $\Raa$, defined as the ratio between the yields measured 
in heavy-ion and $\pp$ collisions after normalizing the A--A yield to the 
average number of nucleon-nucleon collisions for the considered centrality 
class, $\av{\Ncoll}$.
It is anticipated that the medium created in the collision affects the 
abundances and spectra of the originally produced hard probes, resulting in a
break-down of the binary scaling and in a value of $\Raa$ different from 1.
It has however to be considered that other effects related to the presence of 
nuclei in the initial state (e.g. nuclear modifications of the PDFs, Cronin 
enhancement) can break the expected binary scaling.

\subsection{High momentum particle suppression}

Partons are expected to lose energy while traversing the strongly interacting 
medium, via gluon radiation and elastic collisions with the partonic 
constituents.
The measurement of the single-particle nuclear modification factor as a 
function of $\pt$ is the simplest observable sensitive to the energy lost by 
hard partons produced at the initial stage of the collision.
The amount of energy lost is sensitive to the medium properties (density)
and depends also on the path-length of the parton in the deconfined matter
as well as on the properties of the parton probing the medium.

The $R_{AA}$ of unidentified charged particles has been measured by ALICE
up to $\pt=50~\GeV/c$ for various centrality classes~\cite{Otwin}.
For all collision centralities, the $\Raa$ presents a minimum at 
$\pt\approx6-7~\gev/c$ and then increases slowly up to about 30~$\gev/c$. 
A hint of flattening of the nuclear modification factor is observed for 
$\pt>30~\GeV/c$.
The amount of suppression increases with increasing centrality.
For the 5\% most central collisions, the $\Raa$ measured at the LHC is smaller 
than that at RHIC, suggesting a larger energy loss and indicating that the 
density of the medium created in the collision increases with the increase 
of $\sqrts$.
It should also be considered that the fraction of hadrons originating from 
gluon jets increases with increasing centre-of-mass energy and, in radiative 
energy loss models, this is expected to give rise to a lower $\Raa$ because 
gluons lose more energy than quarks while traversing the QGP.

The $\Raa$ was also measured for identified hadrons:
$\pi^{\pm}$ in the range of dE/dx relativistic rise ($3<\pt<20~\gev/c$),
$\pi^{0}$ reconstructed via conversions of the decay photons, 
$K^0_S$ and $\Lambda$~\cite{Schuchmann}.
At high transverse momenta ($\pt>6~\GeV/c$), the suppression of $\pi^{\pm}$,
$K^0_S$ and $\Lambda$ is found to be compatible with that of 
unidentified charged hadrons.
At lower transverse momenta, the charged and neutral pions result to be 
slightly more suppressed (= lower $\Raa$) than charged hadrons and 
the $\Raa$ of $\Lambda$ is significantly larger than that of $K^0_S$ and 
charged hadrons.
This is a consequence of the baryon enhancement observed
in heavy-ion collisions at intermediate momenta, that was also seen
in the $\Lambda$/$K^0_S$ ratio discussed in section~\ref{sec:strang}.
In particular, the $\Lambda$ nuclear modification factor measured at the LHC is
lower than the one observed at RHIC.
This is due to the fact that the different
physical mechanisms that contribute to the $\Raa$ of $\Lambda$ baryons, namely
the baryon enhancement in AA collisions, the canonical suppression in 
the $\pp$ reference and the in-medium energy loss, are quantitatively 
different at the two energies.

\subsection{Open heavy flavours}

Further insight into the energy loss mechanisms can be obtained by measuring the
$\Raa$ for heavy-flavoured hadrons.
Radiative energy loss models predict that quarks lose less
energy than gluons (that have a larger colour charge) and that the amount of
radiated energy decreases with increasing quark mass.
Hence, a hierarchy in the values of the nuclear modification factor is 
anticipated: the $\Raa$ of B mesons should be larger than that of
D mesons that should in turn be larger than that of light-flavour hadrons
(e.g. pions), which mostly originate from gluon fragmentation.

ALICE measured open charm and open beauty
with three different techniques: exclusive reconstruction of $\Dzero$,
$\Dplus$ and $D^{*+}$ hadronic decays at mid-rapidity, 
single electrons after subtraction of a cocktail of background sources at 
mid-rapidity, and single muons at 
forward rapidity~\cite{Castillo}.
The $\Raa$ of prompt D mesons for the 20\% more central collisions shows a 
strong suppression, reaching a factor 4-5 for $\pt > 5~\gev/c$~\cite{Grelli}.
At high $\pt$ the suppression is similar to the one observed for charged 
pions, while at low $\pt$ there seems to be an indication for 
$\Raa$(D)$~>~\Raa$($\pi^\pm$).
The measurement of cocktail-subtracted electrons show, for the 10\% most
central events, a suppression by a factor 1.2-5 in the $\pt$ range between 
3.5 and 6~$\GeV/c$, where charm and beauty decays dominate~\cite{Bailhache}.
For both D mesons and electrons, the suppression is seen to increase with 
increasing centrality.
At forward rapidity, the ratio of central-to-peripheral yield ($\Rcp$)
was measured for muons with $\pt>6~\GeV/c$ where the background contamination
is negligible with respect to muons from heavy flavor decays~\cite{Lopez}.
A suppression of the muon yield which increases with increasing centrality
is observed.

The elliptic flow of $\Dzero$ mesons was also measured and show a hint
of non-zero $v_2$ of charmed hadrons in the range 
$2<\pt<5~\gev/c$~\cite{Bianchin}. 

\subsection{Quarkonia}

Quarkonium states are expected to be suppressed ($\Raa<1$)
in the QGP, due to the color screening of the force which binds the $c\bar{c}$
(or $b\bar{b}$) state.
Quarkonium suppression is anticipated to occur sequentially 
according to the binding energy of each meson: strongly bound states 
($\jpsi$ and $\Upsilon$(1S)) should melt at higher temperatures 
with respect to more loosely bound states.
For collisions at high $\sqrts$, it is also predicted that 
the more abundant production of charm in the initial state would lead
to charmonium regeneration from recombination of $c$ and $\bar{c}$ quarks
at the hadronization, resulting in an enhancement in the number of observed 
$\jpsi$.

ALICE measured the $\jpsi$ nuclear modification factor as a function
of collision centrality at forward rapidity.
$\jpsi$ mesons are measured down to $\pt=0$ without subtracting the
contribution from feed-down from B meson decays.
The resulting $\Raa$ shows a suppression almost independent of centrality and
smaller than that observed by the PHENIX experiment at RHIC in the forward 
rapidity region~\cite{Castillo,Yang}.
At the LHC, ATLAS and CMS measured $\jpsi$ at mid-rapidity and
high $\pt$ ($>6.5~\gev/c$) finding a stronger suppression than that 
observed by ALICE at forward rapidity and low $\pt$, and also stronger
than that measured at RHIC at central rapidity.
Overall, the LHC results on $\jpsi$ nuclear modification factor suggest that 
the $\jpsi$ suppression depends on $\pt$ and that regeneration mechanisms
may play an important role at low $\pt$.
For a deeper understanding, it is crucial to address the initial state effects
by measuring $\jpsi$ production in p--A collisions.

\section{Summary and conclusions}

The first $\PbPb$ run at the LHC enabled the study of heavy-ion physics 
in a new energy regime, about 14 times higher than that attained at RHIC.
The studies on the bulk of soft particles produced in the collision 
demonstrate that the fireball formed in heavy-ion collisions at the LHC 
reaches higher
temperatures and energy densities, lives longer, and expands faster reaching
a larger size at the freeze-out as compared to lower energies.
%In general, the results from soft physics observables show a smooth evolution 
%from RHIC to LHC.
With the first heavy-ion run, we also started to exploit the abundance of 
high $\pt$ and large mass probes which allows high precision measurements
in the hard physics sector.
Further progress is expected from the analysis of the larger $\PbPb$ data 
sample that will be collected in 2011 as well as from the first p--A collisions 
foreseen in 2012.

\end{document}